\documentclass[11pt,twoside]{article}
\usepackage{FUSE2004}
\usepackage{natbib}

\usepackage{epsf}
\usepackage{psfig}
\usepackage{lscape}

\begin{document}
\title{Starbursts in the Far-Ultraviolet}
\author{Timothy M. Heckman} 
\affil{Center for Astrophysical Sciences, Department of Physics \& Astronomy,
Johns Hopkins University}

\begin{abstract}
Starbursts are a significant component of the present-day universe, and
offer unique laboratories for studying the processes that have regulated
the formation and evolution of galaxies and the intergalactic medium.
The combination of large aperture size, medium-to-high spectral resolution,
and access to the feature-rich far-ultraviolet band make FUSE a uniquely
valuable tool for studying starbursts. In this paper, I summarize several
of FUSE's ``greatest hits'' for starbursts.
FUSE observations
of the strong interstellar absorption lines show that powerful starbursts
drive bulk outflows of the neutral, warm, and coronal phases of the
ISM with velocities of several hundred km/s. These
are similar to the outflows seen in Lyman Break Galaxies at high redshift.
The weakness of OVI emission associated with these flows implies
that radiative cooling by coronal gas is not energetically significant.
This increases the likelihood that the flows can eventually escape
the galaxies and heat and enrich the intergalactic medium. FUSE observations
show that local starburst galaxies are quite opaque below the Lyman edge,
with typically no more than $\sim$6\% of the ionizing photons escaping
(even in starbursts with strong galactic winds). This has potentially important
implications for the role of star forming galaxies in the early reionization
of the universe. FUSE observations of molecular hydrogen in starbursts
show very low molecular gas fractions in the translucent ISM
(even in starbursts where mm-wave data show that the ISM is primarily
molecular). This implies that the molecular gas is all in dense clouds
that are completely opaque in the far-UV. The high far-UV
intensity in starbursts photodissociates molecular hydrogen in the diffuse ISM.
Finally, FUSE observations of chemical abundances in the neutral ISM
in dwarf starburst galaxies suggest that the metallicity in the HI (which
dominates the baryonic mass budget) may be systematically lower than in
the HII regions. This would have important implications for
galactic chemical evolution.
\end{abstract}

\section{Introduction}

Starbursts are intense episodes of star formation that dominate
their ``host'' galaxy.
Classically,
they been defined in terms of their short duration. That is,
given the observed star formation rate, the time it
would take to consume the presently available reservoir of
interstellar gas and/or the time it would take to produce the
present-day stellar mass is much less than the age of the universe.
An alternative definition is that a starburst has a high intensity:
the star formation rate per unit area is very large compared
to normal galaxies. As shown by Kennicutt (1998), these two
definitions are functionally equivalent: the gas consumption
time is a systematically decreasing function of the star formation rate
per unit area. Extreme starbursts have gas consumption times
of only $\sim10^8$ years and star formation rates per unit area
thousands of times larger than the disk of the Milky Way.

Starbursts are an important source of light, metals, and high-mass 
star-formation in the local universe (e.g. Heckman 1998). 
Their cosmological relevance
has been highlighted by their many similarities to
star forming galaxies at high-redshift. In particular,
local UV-bright starbursts appear to be good analogs to the
Lyman Break Galaxies (Meurer et al. 1997; Shapley et al. 2003;
Heckman et al. 2005)
and so can be used as a ``training set''
to provide a thorough understanding of rest-frame UV spectral diagnostics
that are critical for studying star-formation in the early universe.
Starbursts can contain millions of OB stars, and hence they
also offer a unique opportunity to test theories
of the evolution of massive stars.

The far ultraviolet spectral window is a key one for understanding
starbursts. This is where the intrinsic spectral energy distribution
of a starburst stellar population peaks. These hot massive stars have
photospheric and wind spectral features in the far-UV that provide key
information about the starburst age, metallicity, and initial
mass function (Robert et al. 2003; Pellerin, this conference).
The far-UV region provides powerful (and in some
cases, unique) diagnostics of the physical, chemical, and
dynamical properties of the interstellar medium from cold molecular
through hot coronal phases. Finally, FUSE is well suited to the problem:
its LWRS aperture is an excellent match to the angular sizes
of the brightest starbursts, and its spectral resolution allows
the interstellar lines to be resolved in almost all cases.

In the following, I will describe what we have learned from FUSE
about the interstellar medium in starbursts.

\section{Dust in Starbursts}

While the peak of the intrinsic spectral energy distribution in starbursts peaks
in the far-UV band, this output is significantly modified by dust. The high 
star formation rate per unit area that defines a starburst directly implies a 
correspondingly high column density of interstellar gas (Kennicutt 1998),
and hence a
high dust column.

The situation is far from hopeless however!
Globally, the volume-averaged emissivity
of star forming galaxies in the local universe in the far-IR (IRAS) and
vacuum UV (GALEX) implies that $\sim$1/3 of the UV starlight escapes and
$\sim$2/3 is absorbed by dust and reradiated in the far-IR (Buat et al. 2005).
However,
this ratio of UV/far-IR flux varies by about three orders-of-magnitude from
galaxy to galaxy, and in a systematic way: the most massive, most metal-rich 
galaxies with the highest star-formation rates are the dustiest
(e.g. Heckman et al.
1998; Martin et al. 2005). Since FUSE is restricted to studying relatively 
UV-bright starbursts, the observed sample is therefore biased in favor of 
metal-poor dwarf starburst systems. However, we have taken pains
in our FUSE program to broadly sample starburst parameter space.

What is the effect of this dust on the emergent far-UV radiation? Meurer, Heckman,
\& Calzetti (1999) used a combination of IUE spectra and IRAS data to demonstrate
a rather surprising result: starburst galaxies show a correlation between
the fraction of the UV radiation that is absorbed and converted to far-IR and
the spectral slope (color) of the emerging vacuum UV radiation. This
result is NOT consistent with a simple picture of a homogeneous mix of stars
and dust in which the observed UV light arises in thin outer ``skin''
(photosphere) of the starburst.
It is instead
consistent with viewing the UV light from the starburst filtered through an ISM filled with dusty clouds (Gordon et al. 1997; Charlot \& Fall 2000). 
This model can also account for
the relatively grey starburst extinction curve - or more properly, the ``effective
attenuation law'' (Calzetti 2001). Leitherer et al. (2002) and 
Buat et al. (2002) have used HUT and FUSE spectra respectively to show that this
effective attenuation law can be smoothly extrapolated from the IUE band into
the far-UV band. 

\section{Starburst-Driven Outflows}

\subsection{Background}

By now, it is well-established that galactic-scale outflows of gas
are a ubiquitous phenomenon in the most actively star-
forming galaxies in the local universe (see Heckman 2002 for a recent review).
These outflows are potentially very important in the evolution of galaxies and
the intergalactic medium. For example, by selectively blowing metals out of 
shallow galactic potential wells, they may explain the tight relation between mass
and metallicity in galaxies (Larson 1974; Tremonti et al. 2004). This same process
would have enriched the intergalactic medium in metals at early times (Adelberger et al. 2003),
and could be responsible for the fact that the majority of metals in galaxy clusters
are in the intracluster medium (e.g. Loewenstein 2004).

The engine that drives the observed outflows in
starbursts is the mechanical energy supplied by massive stars in
the form of supernovae and stellar winds (Leitherer \& Heckman
1995).
The dynamical evolution of a starburst-driven outflow has been
extensively discussed (e.g. Suchkov et al. 1994
Tenorio-Tagle \& Munzo-Tunon 1998; Strickland \& Stevens 2000).
Briefly, the deposition of mechanical energy by
supernovae and stellar winds results in an over-pressured cavity
of hot gas inside the starburst.
This hot gas will expand, sweep
up ambient material and thus develop a bubble-like structure.
If the ambient medium is stratified (like a disk), the superbubble will
expand most rapidly in the direction of the vertical pressure
gradient. After the superbubble size reaches several disk vertical scale
heights, the expansion will accelerate, and it is believed that
Raleigh-Taylor instabilities will then lead to the fragmentation
of the bubble's outer wall (e.g. MacLow, McCray, \& Norman 1989).
This allows the hot gas to ``blow out''
of the disk and into the galactic halo in the form of a weakly
collimated bipolar outflow (i.e. the flow makes a transition from
a superbubble to a superwind). 
The wind will then carry entrained interstellar material out of the galactic disk
and into the halo, and will also accelerate ambient halo clouds.
These outflowing clouds
will give rise to blueshifted interstellar absorption-lines in starbursts.

\subsection{The FUSE Perspective}

FUSE has contributed to our understanding of these outflows in two ways. First,
the interstellar absorption lines in the FUSE band allow us to study the
dynamics
of the neutral, warm, and coronal phases of the outflow. 
Second, observations of OVI line emission with FUSE allows us to assess the 
energetic importance of radiative cooling by the coronal gas. This has important
implications for the dynamical evolution of the wind.

The use of interstellar absorption-lines to study starburst outflows
offer several distinct advantages. Since
the gas is seen in absorption against the background starlight, there is no
possible ambiguity as to the sign (inwards or outwards) of any radial flow
that is detected, and the outflow speed can be measured directly.
Moreover, the strength of the absorption will be related to the
column density of the gas. In contrast, the X-ray or optical surface-brightness
of the emitting gas is proportional to the emission-measure. Thus, the
absorption-lines more fully probe the whole range of gas densities in the
outflow, rather than being strongly weighted in favor of the densest material
(which may contain relatively little mass). Finally, the results of these studies
can be directly compared to the properties of the outflows seen in the interstellar
absorption lines in high-redshift Lyman Break Galaxies (e.g. Shapley et al. 2003).

We (Heckman et al. 2001a; Vasquez et al. 2004) have undertaken FUSE and STIS 
investigations respectively of the dwarf starburst
galaxy NGC 1705. In combination, these serve as a nice case study of the
power of absorption
line spectroscopy to elucidate the physics of starburst driven outflows. 
This nearby (D = 6.2 Mpc) dwarf starburst
was first investigated in detail by Meurer et al. (1992), who
established it as a prototypical example of a dwarf starburst
undergoing mass-loss. They were able to delineate a kpc-scale
fragmented ellipsoidal shell of emission-line gas that was expanding
at roughly 50 km s$^{-1}$ along our line-of-sight. They also showed that the
population of supernovae in the young super star cluster (NGC~1705-1)
was energetically-sufficient to drive this flow. 

Our analysis of FUSE and STIS echelle mode spectra show that the dynamics of
the outflow are quite different in the neutral, warm (photoionized), and
coronal (shock-heated) phases. 
The coronal phase
gas (as probed by OVI, CIV, SiIV, and SIV) is flowing out of the starburst
at a velocity of $\sim$80 km s$^{-1}$. However, the mass and kinetic energy in
the outflow
is dominated by the warm photoionized gas which is also seen through its optical line-emission.
The kinematics of this warm gas are compatible
with a simple model of the expansion at $\sim$50 km/s of a
superbubble
driven by the collective effect of the kinetic energy supplied
by supernovae in the starburst.
However, the observed properties
of the OVI absorption in NGC~1705 are not consistent with
the simple superbubble model, in which the OVI would arise in a conductive
interface inside the superbubble's outer shell.
The relative outflow speed of the OVI
is too high and the observed column density
is much too large.

We argue that the superbubble
has begun to blow out of the ISM of NGC~1705. During this
blow-out phase the superbubble shell accelerates and fragments. The
resulting hydrodynamical interaction as hot outrushing gas flows between
the cool shell fragments will create intermediate-temperature
coronal gas that can produce the observed OVI absorption.
For the observed flow speed,
the observed OVI column density is just what is expected for
gas that has been heated and which then cools radiatively
(Heckman et al. 2002).

\begin{figure}[!ht]
\plotone{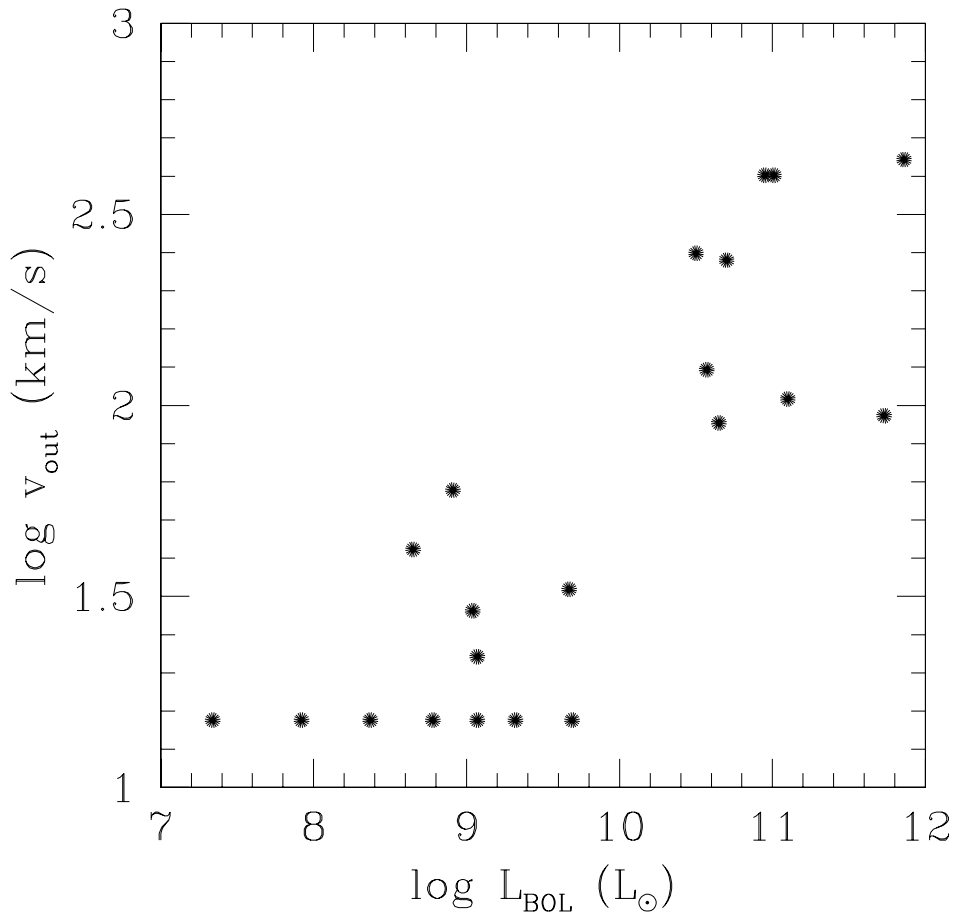}
\caption{The blueshift of the centroid of the interstellar absorption
lines (relative
to the galaxy systemic velocity) is plotted as a function of the starburst
luminosity (taken as the sum of the far-UV and far-IR luminosity) for a
sample of 21 starbursts observed by FUSE. Outflows at velocities of
over a hundred km/s are common in powerful starbursts. The seven data points
plotted at log $v_{out}$ = 1.2 represent upper limits on outflow speed.}
\end{figure}

\begin{figure}[!ht]
\plotone{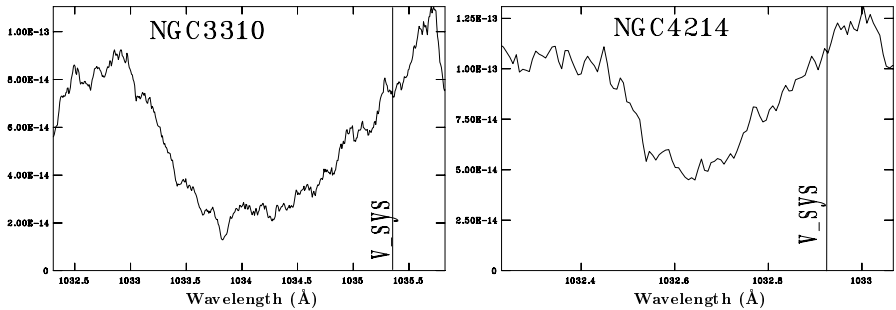}
\caption{$FUSE$ spectra of the OVI$\lambda$1031.9 interstellar absorption-line,
tracing outflowing coronal-phase gas. The absorption covers the range
from $v_{sys}$ to a maximum blueshift of $\sim$700 km s$^{-1}$ in
the powerful starburst NGC 3310 (left panel) and $\sim$140 km s$^{-1}$
in the starbursting irregular galaxy NGC 4214 (right panel).}
\end{figure}

In addition to detailed investigation of favorable cases like NGC~1705, we
are also undertaking a comprehensive survey of the dynamics of the ISM
in a large sample of starburst galaxies observed to date with FUSE. In a first 
pass through the data we have identified in each starburst the strongest interstellar
lines that are relatively ``clean'' (not severely blended with foreground
Milky Way absorption lines). These are typically lines arising in the neutral
or warm-ionized phases (we exclude OVI in this first pass, since it lies in a
complex portion of the spectrum). We then have measured the line width and
radial velocity relative to the published galaxy systemic velocity. The first
results are shown in Figure 1. It is clear that there is a strong trend for the
outflow velocity to increase with the intrinsic UV luminosity of the starburst
(which will be proportional to the star formation rate, and hence to the supernova
heating rate). For starbursts more luminous than a few $\times 10^{10} L_{\odot}$
(SFR $> 10 M_{\odot}$ per year), the outflow velocity at line center is several
hundred km/s. The lines are very broad, and the maximum outflow speeds approach
$\sim10^3$ km/s (see Figure 2). This is consistent with a simple model in which
the clouds that are initially at rest are accelerated to some terminal
velocity by the ram pressure of the wind (see Heckman 2002).

The impact that these outflows have on the evolution
of galaxies and the intergalactic medium depend critically upon whether they
are able to escape the galaxy potential well and eject material into the intergalactic medium.
A necessary condition for the outflow to escape is that radiative losses
are not severe enough to drain energy from the wind, causing it to stall
(e.g. Wang 1995). The X-ray luminosity of the wind is typically on
of-order 1\% of the rate at which supernovae supply kinetic energy.
Thus, radiative losses from hot ($T \geq 10^6$ K) gas will not be
dynamically significant. FUSE observations of OVI emission allow us to measure
the importance of cooling by coronal gas. So far, OVI emission has been searched
for in four starbursts: M~82 (Hoopes et al. 2003), 
NGC~1705 (Heckman et al. 2001a), NGC~3079
(Hoopes, this conference), and NGC~4631 (Otte et al. 2003).
The line is only detected in the last case,
and in all four galaxies the data imply that 
radiative cooling by the coronal gas is not sufficient
to quench the outflow.

\section{The Escape of Ionizing Radiation}

The intergalactic medium (IGM) contains the bulk of the baryons in the
universe (e.g., Fukugita, Hogan, \& Peebles 1998).
Determining the source and strength of the metagalactic ionizing radiation field
and documenting its cosmic evolution is crucial to understanding
the fundamental properties of the IGM at both low- and high-redshift.
The two prime candidates for producing the background are QSOs
and star-forming galaxies. While the contribution to the ionizing background
from QSOs can be estimated with reasonable accuracy, considerably
less is known about the contribution from galaxies. QSOs
alone appear inadequate to produce the inferred background
(e.g, Madau, Haardt, \& Rees 1999), especially at $z \geq$ 3
where their co-moving space density declines
steeply with increasing redshift (Fan et al. 2001).
Moreover, there are only rather
indirect constraints on the contribution of galaxies to the ionizing
background in the low-redshift universe (e.g.,
Giallongo, Fontana, \& Madau 1997; Devriendt et al. 1998; Shull et al. 1999).

The greatest uncertainty in determining the role of star forming galaxies to the
reionization of the universe is the value for $f_{esc}$ - the fraction of the ionizing
photons that escape and reach the IGM. It seems clear that the leakage
of ionizing radiation out of galaxies must be determined by the
structure/topology of the ISM. The hope is that investigations of local star forming
galaxies will allow us to understand the physical processes that determine $f_{esc}$
so that we can apply these lessons to high redshift galaxies for which our information
is less complete.

Leitherer et al. (1995) reported the first direct measurements of
$f_{esc}$ using the Hopkins Ultraviolet Telescope to observe
below the rest-frame Lyman edge in a sample of four local starbursts,
and these data were later reanalyzed by
Hurwitz, Jelinsky, \& Dixon (1997). The
resulting upper limits on $f_{esc}$ were typically 10\%.
Deharveng et al. (2001) have obtained similar data with FUSE for the
starburst galaxy Mrk~54 at z = 0.0448. No flux was detected below the Lyman edge in the rest frame.
By comparison with the number of ionizing photons derived from the H$\alpha$
line, they set an upper limit to $f_{esc}$ of 6\%. Similar types of investigations
have been undertaken at high redshift with conflicting results (Steidel et al.
2001; Malkan et al. 2003).

My colleagues and I have used FUSE in a different way to constrain $f_{esc}$ 
in a sample five of the UV-brightest local starburst galaxies (Heckman et al. 2001b).
We showed that the strong CII$\lambda$1036 interstellar absorption-line is black in
its core. Since the photoelectric opacity of the neutral ISM
below the Lyman-edge will be significantly larger than in the CII line,
we were able to use these data to set a typical upper limit
on $f_{esc}$ of 6\% in these galaxies. Inclusion of absorption of Lyman
continuum photons by dust grains
will further decrease $f_{esc}$ (by up to an order of magnitude in some cases).
We also assessed the idea that
the strong galactic winds discussed above can clear channels
through their neutral ISM and increase $f_{esc}$
(e.g. Fujita et al. 2003). We showed empirically that such
outflows may be a necessary - but not sufficient - part of the
process for creating
a relatively porous ISM.

\section{Molecular Gas}

Molecular hydrogen is the fuel for star formation, so it is natural to expect large amounts of $H_2$
to exist in starbursts. Indeed, millimeter-wave observations using CO as a
tracer imply that molecular gas is the dominant phase by mass in typical
starbursts (e.g Gao \& Solomon 2004).
However, observations of some dwarf starburst galaxies reveal little or no CO
emission, such as NGC 1705 (Greve et al. 1996) and I Zw 18 (e.g., Gondhalekar et al. 1998). 
The lack of CO detections is difficult to interpret for metal-poor galaxies because the CO
to $H_2$ conversion is metallicity dependent (Wilson 1995). 
There are numerous transitions of molecular hydrogen in the far-UV.
Furthermore, far-UV absorption studies can probe $H_2$ to column densities much lower than can be
studied through CO mm-wave line emission,
making it possible to study $H_2$ in the diffuse interstellar medium (ISM).
However, unlike radio observations, far-UV measurements are profoundly affected by extinction. 

We have used FUSE to search for $H_2$ absorption in five starburst galaxies: NGC 1705, NGC 3310, NGC 4214, 
M83 (NGC 5236), and NGC 5253 (Hoopes et al. 2004). We tentatively detected
weak absorption in M83 and NGC 5253 and set upper limits in the other galaxies. Conservative upper limits
on the mass of molecular gas detected
with FUSE are many orders of magnitude lower than those inferred from CO mm-wave measurements for the four
galaxies in our sample in which CO has been detected. 

This indicates that almost all the $H_2$ in starbursts is in the form of clouds with column densities
high enough to make them
completely opaque to far-UV light. This gas can therefore can not be probed with far-UV absorption
measurements. The far-UV continuum visible in the FUSE spectra passes through the translucent ISM
between the dense molecular clouds, which must then have an areal covering factor less than one.
The complex observational biases
related to varying extinction across the extended UV emission in the FUSE apertures prevent an unambiguous
characterization of the diffuse $H_2$ in these starbursts. However, the
evidence suggests that there is a significantly lower molecular fraction in the diffuse interstellar medium
compared to similarly reddened sight lines in the Milky Way.

This is consistent with the higher photo-destruction
rate of $H_2$ due the greatly elevated intensity of the far-UV radiation field in the diffuse ISM
in starbursts compared to the Milky Way (Vidal-Madjar et al. 2000).
It is also consistent with qualitatively similar results
in the Magellanic Clouds (Tumlinson et al. 2002; Tumlinson, this meeting).

\section{The Metallicity of the Neutral Phase}

The chemical abundances
in galaxies provide a unique probe of the past history of star-formation.
For star-forming galaxies essentially all our information
about chemical abundances pertains to the bright H II regions associated
with recent star-formation. This gas is a very minor fraction of
the total ISM mass and may not be representative (since it is potentially
subject to self-pollution by the associated massive stars). This
problem is especially acute for dwarf galaxies, where the majority
of the total baryonic mass (including stars)
is in the H I phase of the ISM. In particular,
there is considerable
debate about whether metal-poor dwarf starbursts are undergoing their
first episodes of star-formation, making them local examples of
primeval galaxies (e.g., Izotov \& Thuan 1999). While the Oxygen
abundances in the H II regions are typically of-order 1/10 solar in these
galaxies, the abundances in the mass-dominating H I phase could
potentially
be much lower if self-pollution of the H II regions is important.

FUSE spectra allow these ideas to be tested by providing access to
H I and metal absorption lines, which can be used to determine the
abundances in the neutral gas outside of star forming regions. This technique has
been applied to a handful of dwarf starburst galaxies, with the 
results described in a series of papers by several different groups 
(Aloisi et al. 2003; Cannon et al. 2004; Heckman et al. 2001; Lebouteiller et al. 2003; 
Lecavelier des Etangs 2003; Thuan et al. 2002).
For the most part, these investigations have found that the
metal abundances in the neutral gas are significantly lower than those in
H II regions (by factors of $\sim$ 3 to 10). However, uncertainties associated
with optical depth effects and ionization corrections must be borne in mind.
These issues have been discussed in more detail in the contributions by
Aloisi, Cannon, and Lebouteiller at this meeting.

A robust conclusion is that even if the metal abundances from FUSE are taken
as lower limits, they definitely require that substantial previous star formation 
has occurred, and over time scales long enough for the enrichment of the
ISM by intermediate mass stars -in the case of nitrogen- and Type Ia
supernovae -in the case of iron (Aloisi et al. 2003). This
indicates that these dwarf starbursts are not
primeval galaxies undergoing their first significant episode of star formation.

\section{Summary and Future Prospects}

Starbursts are important components of the present day universe,
and wonderful laboratories for studying galaxy evolution, massive stars,
and the ISM. The far-UV band is rich with spectral features that provide
unique diagnostics of the physical, chemical, and dynamical state of
the ISM from its molecular to its coronal phases. The combination of
relatively high spectral resolution and a large spectroscopic aperture
make FUSE very well suited to the investigation of starbursts.
In this review I have
summarized some of the highlights of these investigations:

\begin{itemize}
\item
Powerful starbursts drive bulk outflows of the ISM into their galaxy halo
at velocities of several hundred km/s. Similar outflows are seen
in Lyman Break Galaxies at high-redshift.
\item
Radiative cooling/quenching of these outflows by coronal gas
is not dynamically significant. This supports the idea that
they are the mechanism
by which metals were ejected from low mass galaxies and the IGM was
metal-enriched.
\item
Present-day starbursts are quite opaque to their Lyman continuum
radiation. This has interesting implications for the possible reionization
of the universe by early starbursts.
\item
The translucent ISM in starbursts has a very low molecular gas
fraction (most likely due to a high ambient UV intensity in the ISM).
\item
The neutral ISM in dwarf starbursts appears to be significantly
less metal-enriched than the HII regions. If confirmed, this would
have important implications for the chemical evolution of galaxies,
since the neutral phase of the ISM dominates the baryonic mass-budget
in dwarf galaxies like these.
\end{itemize}

The future prospects are very bright. The All-sky Imaging Survey
of the GALEX mission
(Martin  et al. 2005) can provide a large sample of
starbursts in the local universe that are bright enough for FUSE to
obtain very high quality spectra with moderate exposures times. This
offers us the opportunity to attack the problems summarized above
for a sample large enough to draw statistically robust conclusions
across the broad range of fundamental starburst properties (mass,
metallicity, star formation rate, etc).

\acknowledgements 
It is a great pleasure to thank the whole FUSE team for enabling
all the science discussed above. I would also like to thank my collaborators
on these FUSE programs (A. Aloisi, D. Calzetti, C. Hoopes, J.C. Howk,
C. Leitherer, A. Pellerin, C.L. Martin, G. Meurer, C. Robert, S. Savaglio,
K. Sembach, and D. Strickland). This
work was supported in part by NASA FUSE and LTSA grants to me and my
collaborators.

% If you wish to use BiBTeX uncomment and fill in the .bib file name.  Note that 
% we are still (July 23) waiting for input from the ASP as to which 
% "bibliographystyle" to use.  "natbib" is unlikely to be the right one, but is 
% left here as a place holder.

%\bibliography{bib-file}

\begin{thebibliography}{}
\bibitem[]{}Adelberger, K., Steidel, C., Shapley, A., \& Pettini, M. 2003, \apj, 584, 45
\bibitem[]{}Aloisi, A., Savaglio, S., Heckman, T., Hoopes, C., Leitherer, C., \& Sembach,
K. 2003, \apj, 595, 760
\bibitem[]{}Buat, V., Burgarella, D., Deharveng, J. M., \& Kunth, D. 2002, \aap, 393, 33
\bibitem[]{}Buat, V. and the GALEX Science Team 2005, \apj, in press
\bibitem[]{}Calzetti, D. 2001, \pasp, 113, 1449
\bibitem[]{}Cannon, J., Skillman, E., Sembach, K., \& Bomans, D. 2004, \apj, in press
\bibitem[]{}Deharveng, J.-M., Buat, V., Le Brun, V., Milliard, B., Kunth, D., Shull, J. M.,
\& Gry, C. 2001, \aap, 375, 805
\bibitem[]{}
Devriendt, J., Sethi, S., Guiderdoni, B., \& Nath, B. 1998, \mnras,
298, 708
\bibitem[]{}Fan, X. et al. 2001, \aj, 122, 2833
\bibitem[]{}Fujita, A., Martin, C. L., Mac Low, M.-M., \& Abel, T. 2003, \apj,
599, 50
\bibitem[]{}Fukugita, M., Hogan, C., \& Peebles, P. J. 1998, \apj, 502, 518
\bibitem[]{}Gao, Y., \&  Solomon, P. 2004, \apj, 606, 271
\bibitem[]{}Giallongo, E., Fontana, A., \& Madau, P. 1997, \mnras, 289, 629
\bibitem[]{}Gondhalekar, P., Johansson, L., Brosch, N., Glass, I., \& Brinks, E. 1998,
\aap, 335, 152
\bibitem[]{}Gordon, K,, Calzetti, D., \& Witt, A. 1997, \apj, 487, 625
\bibitem[]{}Greve, A., Becker R., Johansson L., \& McKeith C. 1996, \aap, 312, 391
\bibitem[]{}Heckman, T. 1998, in ``Origins'', ASP Conference Series, Vol. 148, 1998, ed. C.
Woodward, J. M. Shull, and H. Thronson, Jr., p.127
\bibitem[]{}Heckman, T. 2002, in ``Extragalactic Gas at Low Redshift'', ASP Conference
Proceedings Vol. 254. Edited by J. Mulchaey and J. Stocke, p. 292
\bibitem[]{}Heckman, T., Robert, C., Leitherer, C., Garnett, D., \&
van der Rydt, F. 1998, \apj, 503, 646
\bibitem[]{}Heckman, T., Sembach, K., Meurer, G., Strickland, D., Martin, C.L.,
Calzetti, D., \& Leitherer, C. 2001a, \apj, 554, 1021
\bibitem[]{}Heckman, T., Sembach, K., Meurer, G., Leitherer, C., Calzetti, D.,
\& Martin, C. L. 2001b, \apj, 558, 56
\bibitem[]{}Heckman, T., Norman, C., Strickland, D., \& Sembach, K. 2002,
\apj, 577, 691
\bibitem[]{}Heckman, T., and the GALEX Science Team 2005 \apj, in press
\bibitem[]{}Hoopes, C., Heckman, T., Strickland, D., \& Howk, J. C. 2003, \apj, 596, L175
\bibitem[]{}Hoopes, C., Sembach, K., Heckman, T., Meurer, G., Aloisi, A., Calzetti, D.,
Leitherer, C., \& Martin, C. L. 2004, \apj, 612, 825
\bibitem[]{}Hurwitz, M., Jelinsky, P., \& Dixon, W. 1997, \apj, 481, 31
\bibitem[]{}Izotov, Y., \& Thuan, T. 1999, \apj, 511, 639
\bibitem[]{}Kennicutt, R. 1998, \apj, 498, 541
\bibitem[]{}Larson, R. 1974, \mnras, 169, 229
\bibitem[]{}Lebouteiller, V., Kunth, D., Lequeux, J.,
Lecavelier des Etangs, A., Desert, J.-M., Hebrard, G., \& Vidal-Madjar, A.
2004, \aap, 415, 55
\bibitem[]{}Leitherer, C., \& Heckman, T. 1995, \apjs, 96, 9
\bibitem[]{}Leitherer, C., Ferguson, H., Heckman, T., \& Lowenthal, J. 1995, \apj, 454, L19
\bibitem[]{}Leitherer, C., Li, I.-H., Calzetti, D., \& Heckman, T. 2002, \apjs, 140, 303
\bibitem[]{}Lecavelier des Etangs, A., Desert, J.-M, Kunth, D., Vidal-Madjar, A.,
Callejo, G.,  Ferlet, R., Hebrard, G., \& Lebouteiller, V. 2004, \aap, 413, 131
\bibitem[]{}Loewenstein, M. 2004, in ``Origin and Evolution of the Elements'',
Cambridge University Press, edited by A. McWilliam and M. Rauch, p. 425
\bibitem[]{}MacLow, M., McCray, R., \& Norman, M. 1989, \apj, 337, 141
\bibitem[]{}Madau, P., Haardt, F., \& Rees, M. 1999, \apj, 514, 648
\bibitem[]{}Malkan, M., Webb, W., \& Konopacky, Q. 2003, \apj, 598, 878
\bibitem[]{}Martin, C.D., and the GALEX Science Team 2005 \apj, in press
\bibitem[]{}Meurer, G., Freeman, K., Dopita, M., \& Cacciari, C. 1992,
\aj, 103, 60
\bibitem[]{}Meurer, G., Heckman, T., Leitherer, C., Lowenthal, J.,
\& Lehnert, M. 1997, \aj, 114, 54
\bibitem[]{}Meurer, G., Heckman, T., \& Calzetti, D. 1999, \apj, 521, 64
\bibitem[]{}Otte, B., Murphy, E., Howk, J. C., Wang, Q. D., Oegerle, W., \& Sembach, K.
2003, \apj, 591, 821
\bibitem[]{}Robert, C., Pellerin, A., Aloisi, A., Leitherer, C., Hoopes, C., \&
Heckman, T. 2003, \apjs, 144, 21
\bibitem[]{}Shapley, A., Steidel, C., Pettini, M., \& Adelberger, K. 2003, \apj, 588, 65
\bibitem[]{}Shull, S.M., Roberts, D., Giroux, M., Penton, S., \& Fardal, M.
1999, \aj, 118, 1450
\bibitem[]{}Steidel, C., Pettini, M., \& Adelberger, K. 2001, \apj,
546, 665
\bibitem[]{}Strickland, D., \& Stevens, I. 2000, \mnras, 314, 511
\bibitem[]{}Suchkov, A., Balsara, D., Heckman, T., \& Leitherer, C. 1994,
\apj, 430, 511
\bibitem[]{}Tenorio-Tagle, G., \& Munoz-Tunon, C. 1998, \mnras, 293, 299
\bibitem[]{}Thuan, T., Lecavelier des Etangs, A., \& Izotov, Y. 2002, \apj, 565, 941
\bibitem[]{}Thuan, T., Lecavelier des Etangs, A., \& Izotov, Y. 2002, \apj, 565, 941
\bibitem[]{}Tremonti, C. et al. 2004, \apj, in press
\bibitem[]{}Tumlinson, J., et al. 2002, \apj, 566, 857
\bibitem[]{}Vasquez, G., Leitherer, C., Heckman, T., Lennon, D., de Mello, D., Meurer, G.,
\& Martin, C. L. 2004, \apj, 600, 162
\bibitem[]{}Vidal-Madjar, A., et al. 2000, \apj, 538, L77
\bibitem[]{}Wang, B. 1995, \apj, 444, 590
\bibitem[]{}Wilson, C. 1995, \apj, 448, L97
\end{thebibliography}
%\bibliographystyle{natbib}

% For using the "thebibliography" environment use these.
% See the "Authors Instructions" for details.

\end{document}